\documentclass[a4paper,fleqn,usenatbib]{mnras}

\usepackage{epsfig}
\usepackage{times}
\usepackage{amsmath}

\newif\ifAMStwofonts
\newcommand{\target}{3FGL\,J0212.1+5320}

\newcommand{\vsini}{$v_{\rm rot}\,\rm sin\,{\it i}$}
\newcommand{\req}{$r_{\rm eq}$}
\newcommand{\kms}{\,km\,s$^{-1}$}
\newcommand{\ergs}{\,erg\,s$^{-1}$\,cm$^{-2}$}
\newcommand{\erg}{\,erg\,s$^{-1}$}
\newcommand{\msun}{\,$M_{\sun}$}

\title[The redback pulsar 3FGL\,J0212.1+5320]
{Properties of the redback millisecond pulsar binary 3FGL\,J0212.1+5320}

\author[T.\,Shahbaz et al. ]
       {T.\,Shahbaz,$^{1,2}$\thanks{E-mail: tsh@iac.es}
                M.\,Linares,$^{3,1,2}$
                R.\,P.\,Breton$^4$\\  
$^{ \it 1}$Instituto de Astrof\'\i{}sica de Canarias (IAC), E-38200 La Laguna, 
Tenerife, Spain \\
$^{ \it 2}$Departamento de  Astrof\'\i{}sica, Universidad de La Laguna (ULL), 
E-38206 La Laguna, Tenerife, Spain \\
$^{ \it 3}$Department de F{\'i}sica, EEBE, Universitat Polit{\`e}cnica de 
Catalunya, c/ Eduard Maristany 10, 08019 Barcelona, Spain \\
$^{ \it 4}$Jodrell Bank Centre for Astrophysics, School of Physics and 
Astronomy, The University of Manchester, Manchester M13 9PL, UK 
}

\pagerange{\pageref{firstpage}--\pageref{lastpage}}
\pubyear{2014}

\begin{document} 
\maketitle 
\begin{abstract} 

\noindent
Linares et al. (2016) obtained quasi-simultaneous 
$g'$, $r'$ and $i'$-band light curves and an absorption line radial velocity curve of the 
secondary star in the redback system \target. The light curves showed two maxima
and minima primarily due to the secondary star's ellipsoidal 
modulation, but with unequal maxima and minima.
We fit these light curves and radial velocities 
with our X-ray binary model including either a 
dark solar-type star spot or a hot spot due to 
off-centre heating from an intrabinary shock, to account for the unequal maxima. 
Both models give a radial velocity semi-ampltiude
and rotational broadening that agree with the observations. 
The observed secondary star's effective temperature is best matched 
with the value obtained using the hot spot model,
which gives a  neutron star 
and secondary star mass of $M_{\rm 1}$=1.85\,$^{+0.32}_{-0.26}$ \msun\ and 
$M_{\rm 2}$=0.50\,$^{+0.22}_{-0.19}$ \msun, respectively.  
\end{abstract}

\section{Introduction}

Progenitors of binary millisecond pulsars (MSPs) are believed to be 
recycled dead pulsars in low-mass X-ray binaries (LMXBs). According to 
this scenario the neutron star in the LMXB accretes material and angular 
momentum from its late-type companion star and is thereby spun up for 
billions of years to ultimately evolve into a MSP \citep{Alpar82, 
Bhattacharya91}. Compact binary MSPs with orbital periods less than 1 d are commonly classified as 
either ''black widows'' or ''redbacks'' depending on the mass of the 
companion star, $M_{\rm 2}$. Black widows have relatively low-mass degenerate
companion stars 
$0.02\,M_{\odot} \leq M_{\rm 2} \leq 0.05\,M_{\odot }$, whereas
redbacks have relatively more massive non-degenerate
companions $0.2\,M_{\odot} \leq M_{\rm 2} \leq 0.4\,M_{\odot}$
\citep[see][and references therein]{Roberts13}. 
The evolutionary link between redback and black widow MSPs is still 
uncertain. According to \citet{Chen13}, black widows and redbacks are two 
distinct populations of MSPs with different evaporation efficiencies. On 
the other hand, \citet{Benvenuto14} argue that redbacks with compact 
orbits evolve to black widows, while the ones with longer orbital periods 
evolve to MSP-He white dwarf systems, and that  black widows descend 
from redbacks but not all redbacks become black widows.

The Large Area Telescope (LAT) on the {\it Fermi} Gamma-Ray Space Telescope has 
been successful in uncovering binary MSPs because they are $\gamma$-ray 
emitters similar to young pulsars \citep{Abdo09}. Subsequently, thanks to 
targeted radio and X-ray surveys where {\it Fermi}-LAT 
has localized sources, more than 30 black widow and 14 redback systems 
have been discovered \citep{Hessels11, Roberts13, Crawford13, Linares16}. 
To date  there are three (PSR\,J1023+0038, IGR\,J18245-2452 and XSS\,J12270-4859) 
redback systems that transition between 
accretion-powered LMXB states and rotation-powered radio pulsar states 
clearly indicates the close relationship between LMXBs and radio MSPs and 
has provided further support for the recycling scenario 
\citep{Archibald09, Papitto13, Bassa14}.

The optical light curves of black widow and redback binaries show 
large-amplitude variability due to irradiated ellipsoidal modulation of 
the near-Roche filling secondary star. Compared to the black widows, the 
redbacks, with their larger secondary stars and closer distance are 
relatively bright in the optical, allowing for detailed photometric and 
spectroscopic observations \citep{Breton13, Li14, Schroeder14, Linares16}. 
In the last few years, dynamical photometric and spectroscopic studies of 
binary MSPs have largely proven the neutron star masses in these 
systems are generally heavier than the 1.4\,$M_{\odot}$ canonical value, 
as expected theoretically for the binary evolution of MSPs 
\citep{Chen13,Benvenuto14}.

In this paper we present the results of modelling the optical light and 
radial velocity curves of the redback candidate \target\ discovered by 
\citet{Linares16} and \citet{Li16}. First we briefly describe the X-ray binary model used, 
the fitting procedure and the results where we determine the binary system 
masses. Finally we discuss the impact of these results.

\section{\target}
\label{target}

Recently, both \citet{Linares16} and \citet{Li16} presented the discovery 
of a variable optical counterpart to the unidentified $\gamma$-ray source 
\target, and argued that it is a binary "redback" MSP candidate. 
\citet{Linares16} obtained quasi-simultaneous $g'$, $r'$ and $i'$-band 
light curves as well a low- and high resolution spectroscopy. The optical 
light curves obtained by \citet{Linares16} and \citet{Li16} both show two 
maxima and minima, primarily due to the secondary star's ellipsoidal 
modulation. However, the light curve obtained by \citet{Linares16} shows 
unequal maxima, which is not seen in the light curve obtained by 
\citep{Li16} due to very poor orbital phase coverage. From the combined 
photometry and radial velocities 
\citet{Linares16} determined an orbital period of 0.86955(15)\,d, the same 
orbital period was obtained by \citet{Li16} from their $R$ and $g'$-band 
light curves.

From high resolution spectra taken between orbital phases 0.17 and 0.22, 
\citet{Linares16} found the secondary star to have a spectral type of  F6$\pm$2.
They also found no notable changes in spectral type or colour  across the 
binary orbit, not surprising given the long orbital period and hence weak  X-ray
heating effects. They determined a radial velocity curve using the  H$\alpha$
absorption line and from  a sinusoidal fit to the curve obtained a radial
velocity semi-amplitude of  $K_{\rm 2}$=214.1$\pm$5.0\kms. Finally using high
resolution spectroscopy  they estimate the projected rotational velocity of the
secondary star to  be \vsini=73.2$\pm$1.6\kms.

%%%%%%%%%%%%%%%%%%%%%%%%%%%%%%%%%%%%%%%%%%%%%%%%%%%%%%%%%%%%%%%%%%%%%%%
\begin{figure}
\centering
\includegraphics[width=\columnwidth]{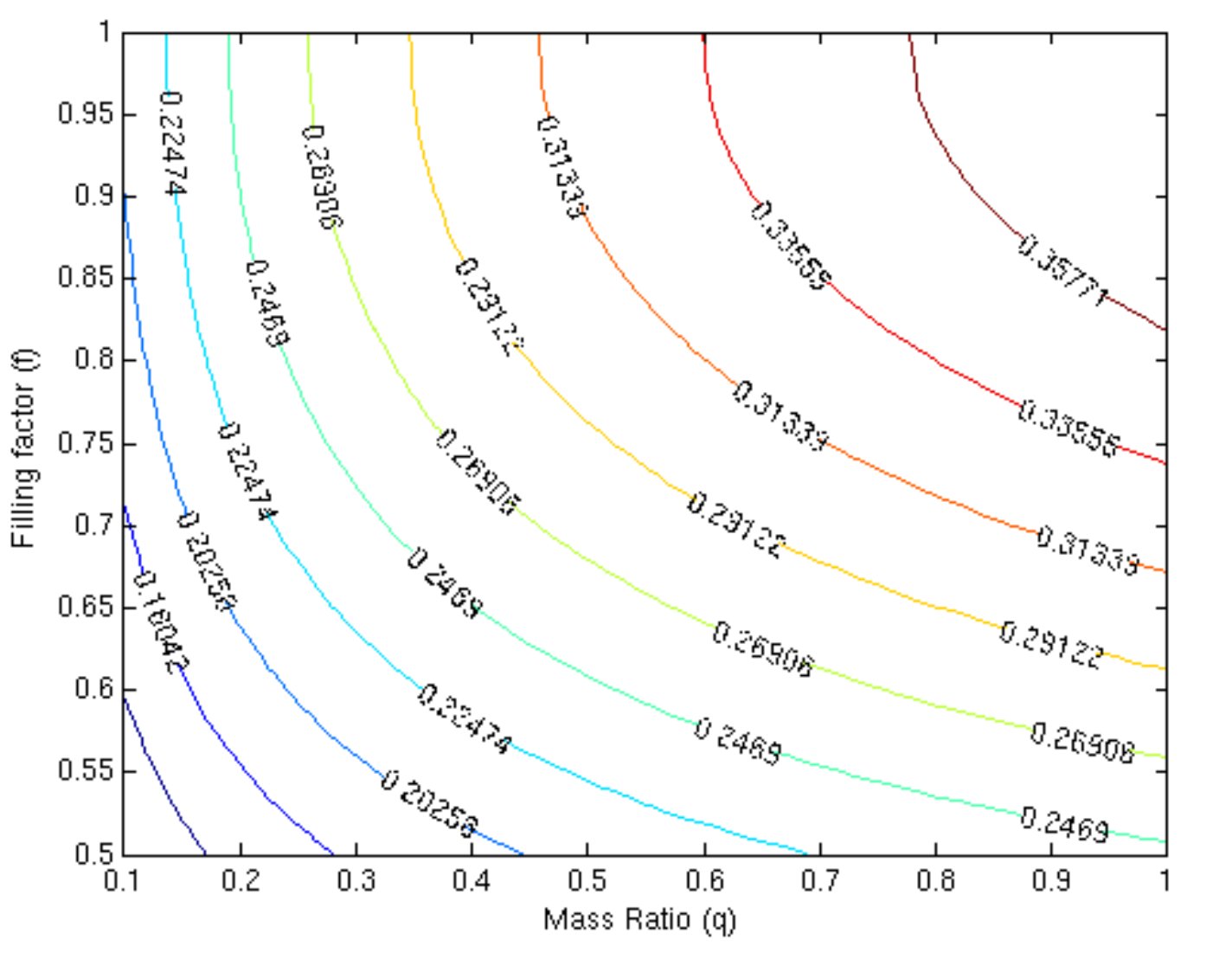}
\caption{Contour plot of the equivalent volume radius  (\req$/a$) for 
different binary mass ratio ($q$) and Roche lobe filling factors ($f$).}
\label{fig:req}
\end{figure}
%%%%%%%%%%%%%%%%%%%%%%%%%%%%%%%%%%%%%%%%%%%%%%%%%%%%%%%%%%%%%%%%%%%%%%%

%%%%%%%%%%%%%%%%%%%%%%%%%%%%%%%%%%%%%%%%%%%%%%%%%%%%%%%%%%%%%%%%%%%%%%%
\begin{figure}
\centering
 \includegraphics[width=\columnwidth, angle=-90, scale=0.6]{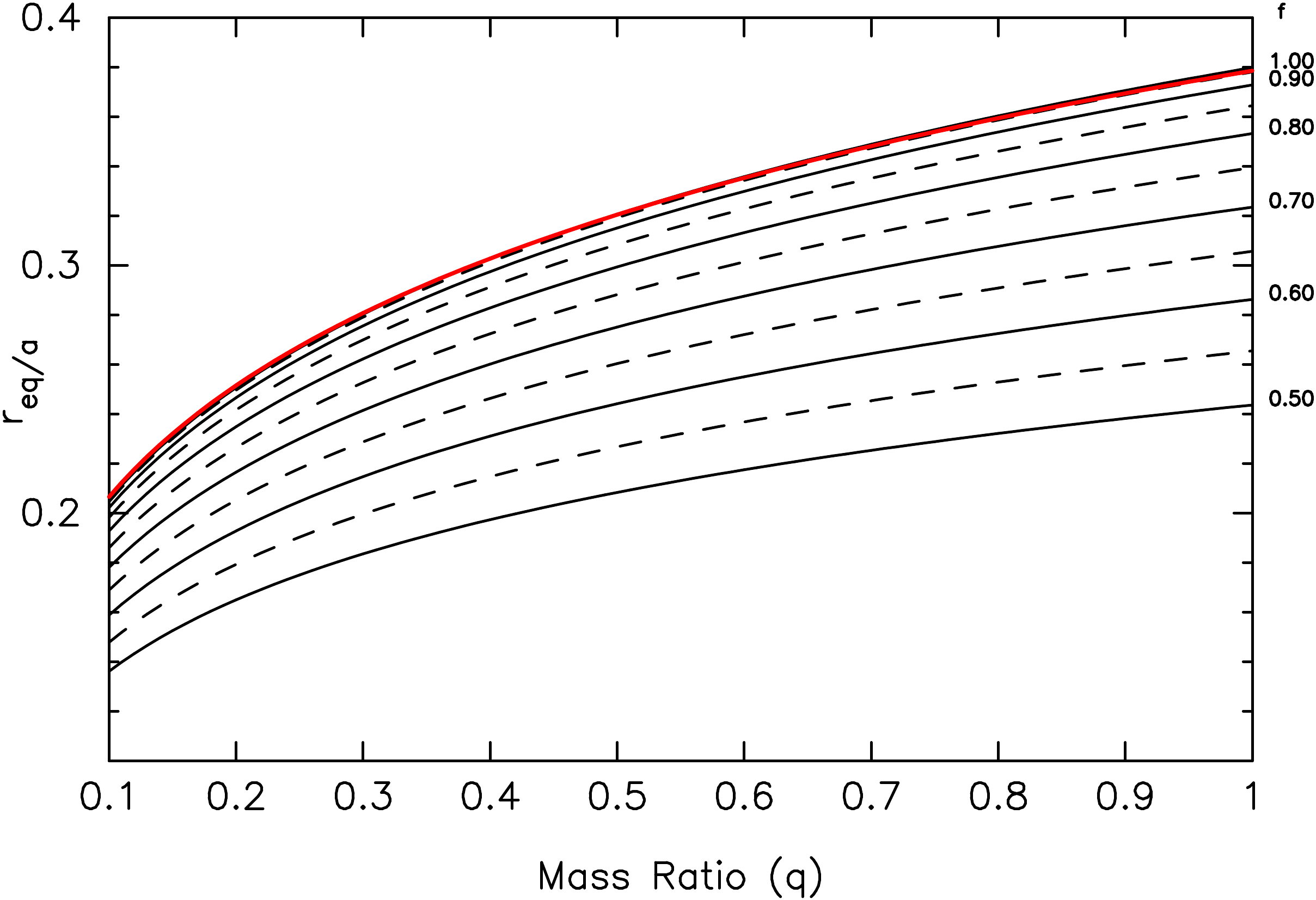}
\caption{The equivalent volume radius radius ($r_{\rm eq}/a$) versus 
binary mass ratio ($q$) for different Roche lobe filling factors ($f$). 
From bottom to top $f$=0.6 to 1.0. For comparison we also show Eggleton's 
relation (red line) which assumes a fully Roche filling star, which us 
equivalent to our model with $f$=1.0. }
\label{fig:q-f}
\end{figure}
%%%%%%%%%%%%%%%%%%%%%%%%%%%%%%%%%%%%%%%%%%%%%%%%%%%%%%%%%%%%%%%%%%%%%%%

\section{The X-ray binary model}
\label{sec:model}

We use the X-ray binary light curve model 
\textsc{xrbcurve} described in \citep[][]{Shahbaz03a}, which  has 
successfully been used to model the light curves and radial velocity 
curves of neutron star and black hole X-ray binaries 
\citep[e.g. see ][]{Shahbaz03a, Shahbaz04}.
to fit and interpret the photometric light curves and the 
H$\alpha$ absorption-line 
radial velocity curve of \target\ presented in 
\citet{Linares16}. 
Briefly, the model consists of a binary system in which the primary  
point-like compact object with mass $M_{\rm 1}$ and the secondary 
is a 
Roche lobe filling star with mass $M_{\rm 2}$. 
which is assumed to 
be in a circular orbit and in synchronous rotation. The binary mass ratio 
$q$ is defined as the ratio $M_{\rm 2}$/$M_{\rm 1}$.
%A grid of quadrilaterals of approximately equal area is set up over the 
%Roche surface of the star.
The binary geometry  is determined by the 
binary masses, the orbital inclination $i$, and the Roche lobe 
filling factor $f$ of the secondary star, defined as the ratio of the 
radius of the star from its centre of mass to the inner Lagrangian point 
to the distance towards the inner Lagrangian point.

\subsection{The equivalent volume radius}
\label{sec:evr}

When the secondary stars in interacting binaries such as cataclysmic 
variables or X-ray binaries are tidally locked and in synchronous rotation, 
for a given orbital period the width of the rotationally-broadened absorption
line profile 
arising from the star scales with the size of its Roche lobe. One can 
show that the star's rotational broadening \vsini\ and equivalent volume 
radius \citep{Eggleton83} $r_{\rm eq}/a$ ($a$ is the binary separation), 
defined as the radius 
of a sphere whose volume is the same as the volume of the secondary star, 
are related through the expression

\begin{equation}
\label{eq:1}
v_{\rm rot}\,\sin\,i / K_{\rm 2} = (1+q) r_{\rm eq}(q)  / a.
\end{equation}

\noindent
For a star that fully fills its Roche lobe, \citet{Eggleton83} numerically 
calculated the Roche lobe volume for different mass ratios and determined 
an analytical expression for $r_{\rm eq}/a$ as a function of $q$ (normally 
referred to as Eggleton's formula). However, for stars that do not fill 
their Roche lobe, one cannot use Eggleton's formula to determine $r_{\rm 
eq}/a$, instead one has to calculate $r_{\rm eq}/a$ numerically for a given 
binary $q$ and $f$ configuration. The star's rotational broadening is then 
given by

\begin{equation}
\label{eq:2}
v_{\rm rot}\,\sin\,i / K_{\rm 2} = (1+q)r_{\rm eq}(f,q)/a.
\end{equation}

\noindent
Using our X-ray binary model we numerically determine $r_{\rm eq}/a$ as a 
function of $f$ and $q$, assuming synchronous rotation. For a given $q$ and 
$f$ configuration we determine the binary Roche potential and thus the 
star's Roche lobe. We sample the star's surface with 18340 quadrilaterals 
of approximately equal area and then perform a numerical integration to 
calculate its volume and hence equivalent radius. Fig.\,\ref{fig:req} 
shows a contour plot of the $r_{\rm eq}/a$ values for different $q$ and $f$ 
values. In Fig.\,\ref{fig:q-f} we show the results for $f$ in the range 0.5 
to 1.0 and $q$ in the range 0.1 to 1.0. The model $f$=1.0 is equivalent to 
Eggleton's relation and we find our model and Eggleton's relation are 
consistent to within 0.1 per cent.

%%%%%%%%%%%%%%%%%%%%%%%%%%%%%%%%%%%%%%%%%%%%%%%%%%%%%%%%%%%%%%%%%%%%%%%
\begin{figure}
\begin{center}  
\includegraphics[width=0.85\linewidth, angle=-90]{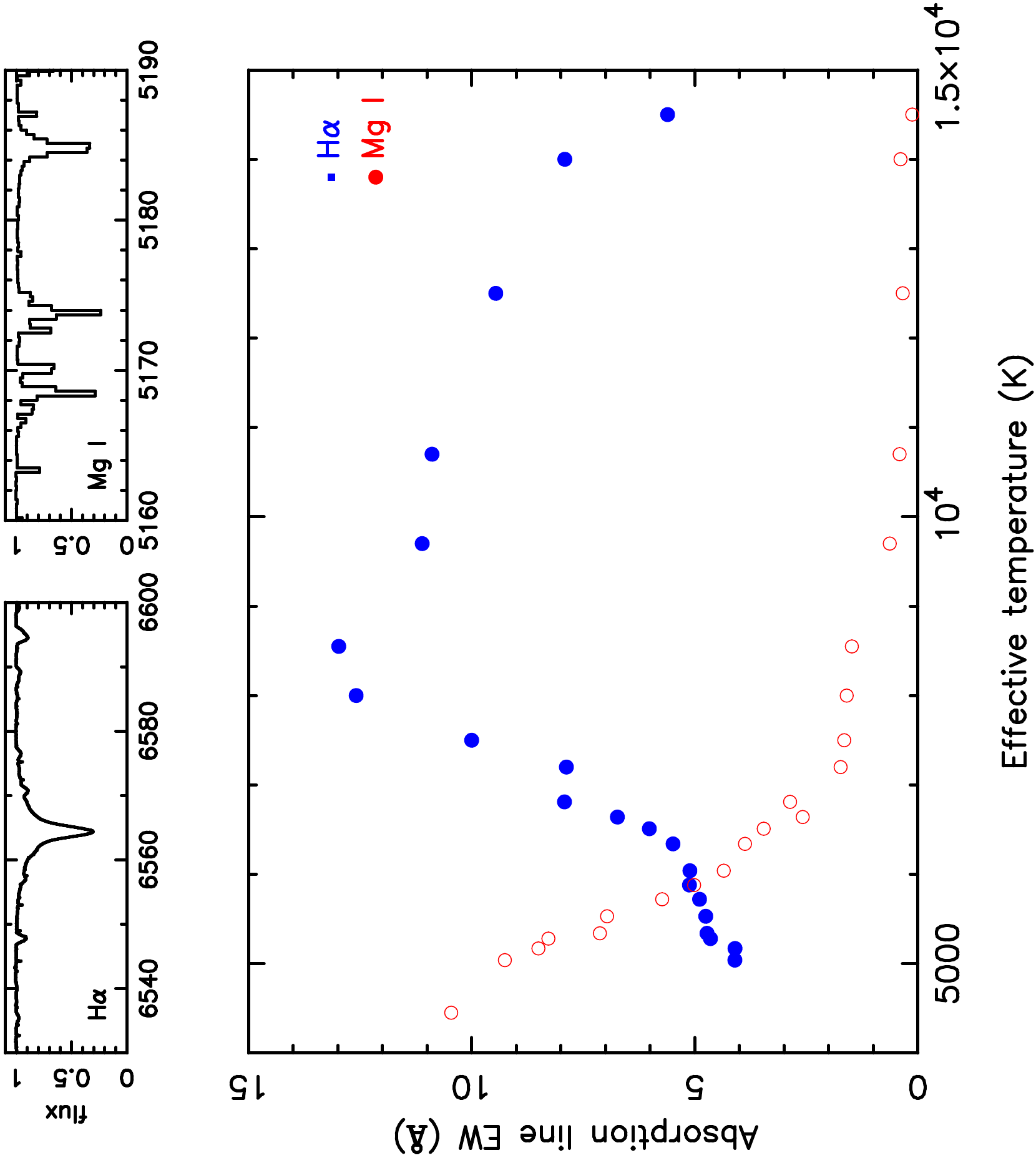}
\end{center}  
\caption{Equivalent width (EW) versus effective temperature ($T_{\rm 
eff}$) relation for H$\alpha$ (blue open  squares) the 
Mg\,\textsc{I}\,b triplet 5167,5172,5183\AA\ (red filled circles) 
are shown. Spectra from the 
VLT-UVES Paranal Observatory Project database  was used
\citep{Bagnulo03}.
Top panel: example spectra showing the 
Mg\,\textsc{I}\,b triplet and H$\alpha$ absorption lines.}
\label{fig:EW}
\end{figure}
%%%%%%%%%%%%%%%%%%%%%%%%%%%%%%%%%%%%%%%%%%%%%%%%%%%%%%%%%%%%%%%%%%%%%%%

\subsection{Light curve}
\label{sec:lcurve}

The secondary star's effective temperature the gravity-darkening exponent 
and reddening given by  $T_{\rm 2}$,  $\beta$ and $E(B-V)$, respectively 
determines the observed light arising from the secondary star. 
The model includes the effects of heating 
of the secondary star by a point source from the 
compact object. 
The irradiating flux $F_{\rm X}$  is thermalised in the 
secondary star's photosphere and is re-radiated locally at a higher 
effective temperature. 
For each point on the star's surface, the effective temperature is calculated 
by combining the intrinsic and incident fluxes. 
We assume the "deep 
heating" approximation where the irradiation 
does not affect the temperature structure of the atmosphere. This means that
 each element radiates as predicted by a 
model atmosphere for a single star. 
The scale of the system is set by 
the distance to the source in parsecs, the orbital period and the radial velocity 
amplitude of the secondary star ($D_{\rm pc}$, $P_{\rm orb}$ and $K_{\rm 2}$, respectively).
We use \textsc{NextGen} model-atmosphere fluxes \citep{Hauschildt99} to 
determine the intensity distribution on the secondary star and 
a quadratic limb-darkening law with coefficients taken from \citet{Claret11}, to correct 
the intensity for limb-darkening.

%%%%%%%%%%%%%%%%%%%%%%%%%%%%%%%%%%%%%%%%%%%%%%%%%%%%%%%%%%%%%%%%%%%%%%%
\begin{figure*}
\includegraphics[width=\linewidth, angle=0]{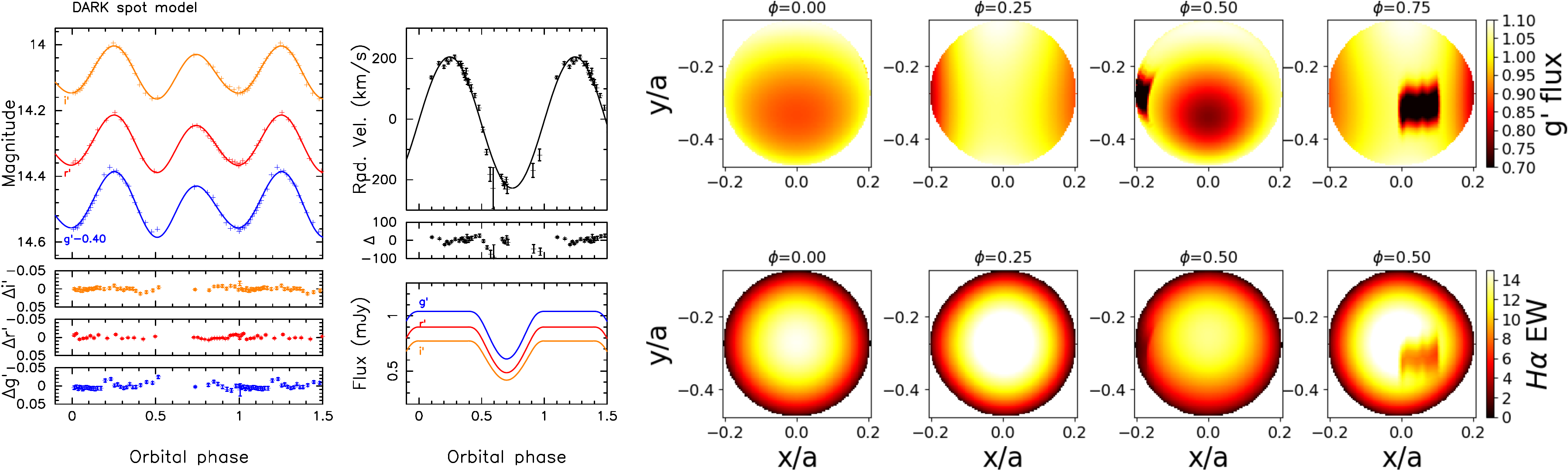} 
\caption{ Left: The results of the simultaneous fits to the light (top 
left) and radial velocity curves (top right) of \target. The $g'$ (blue), 
$r'$ (red) and $i'$ (yellow) light curves and the H$\alpha $absorption line 
radial velocity curve (black) are shown, along with the corresponding best 
fit (solid line) are shown (see Table\,\ref{table:mcmc}). 
Also shown are the corresponding residuals in each band (bottom left), and the 
and the light curve of the additional 
source of light in each band (bottom right). The orbital cycle has been 
repeated for clarity. Right: Projected maps of the observed $g'$ band (top)
H$\alpha$ absorption line strength (bottom) distribution on the secondary 
star at different orbital phases. The best fit dark spot model parameters 
are $f$=0.80, $q$=0.18 and $i$=65\,$^\circ$. The inner Lagrangian point is 
at position (0.0,0.0), (-0.2,0.0), (0.0,0.0) and (0.20,0.0) for orbital 
phases 0.0, 0.25, 0.50 and 0.75, respectively. }
\label{fig:dark}
\end{figure*}
%%%%%%%%%%%%%%%%%%%%%%%%%%%%%%%%%%%%%%%%%%%%%%%%%%%%%%%%%%%%%%%%%%%%%%%

%%%%%%%%%%%%%%%%%%%%%%%%%%%%%%%%%%%%%%%%%%%%%%%%%%%%%%%%%%%%%%%%%%%%%%%
\begin{figure*}
\includegraphics[width=\linewidth, angle=0]{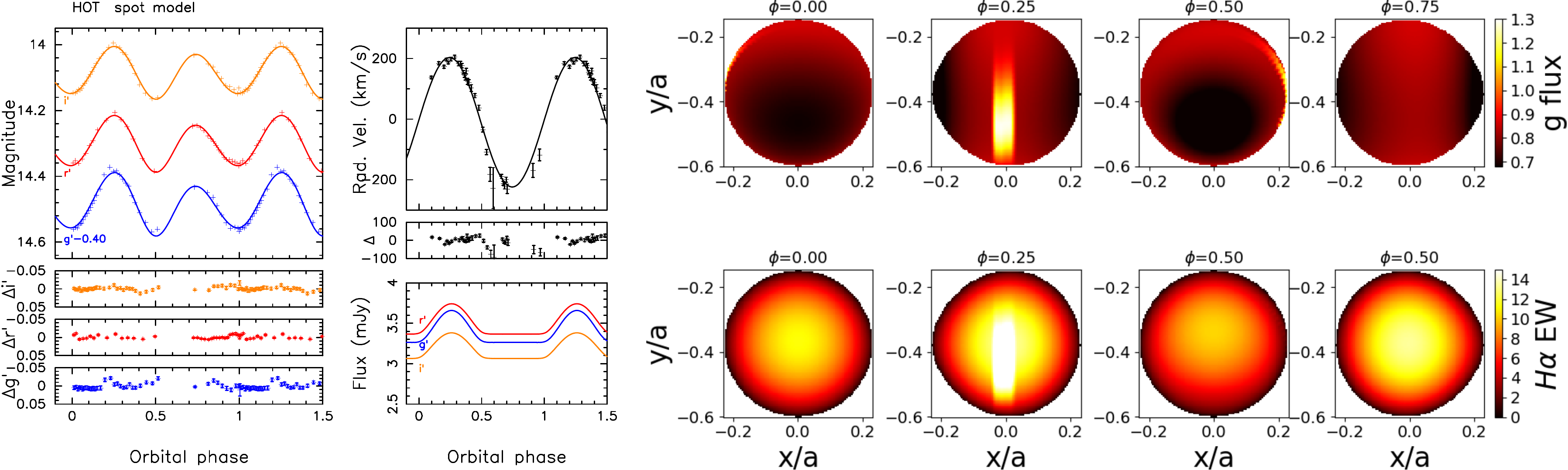}  
\caption{ Same as Fig.\,\ref{fig:dark} but for the hot spot model with 
best fit parameters are $f$=0.76, $q$=0.28 and $i$=69\,$^\circ$. (see 
Table\,\ref{table:mcmc}). }
\label{fig:hot}
\end{figure*}
%%%%%%%%%%%%%%%%%%%%%%%%%%%%%%%%%%%%%%%%%%%%%%%%%%%%%%%%%%%%%%%%%%%%%%%

\subsection{Radial velocity curve}
\label{sec:rcurve}

The optical light curves of binary MSPs show the secondary star's 
ellipsoidal modulation combined with the effects of heating. 
The spectrum shows intrinsic absorption lines arising from the secondary 
star as well as irradiation-induced absorption Balmer lines. Both radial 
velocity curves are distorted because of shift on the centre of mass of 
the lines that are used to determine the radial velocities.

The model light curve is determined by integrating the observed flux from each 
element of area on the star in the observers line-of-sight. For the radial velocity 
curves we specify the strength of the absorption or emission lines over 
the secondary star's surface, and integrate to obtain the corresponding 
line-of-sight radial velocity.  We set the 
absorption line strength given by its equivalent width (EW) according to 
the effective temperature for each element on the star.  However, as 
mentioned in \citet{Phillips99, Shahbaz00} we must also consider the 
consequences of external heating. The vertical temperature gradient in an 
  internally and externally heated atmosphere  produces weaker absorption 
lines than expected from the effective temperature. Since there is no 
decent models to treat the effects of external heating in atmospheres, we 
introduce the factor $F_{\rm AV}$, which represent the fraction of the 
external radiation flux that exceeds the unperturbed flux. A value of $F_{\rm AV}$=1.10 means that if the external radiation flux is greater 
than 10 per cent of the unperturbed flux, then we set the EW for that 
element to zero, otherwise, the absorption line strength takes the EW 
corresponding to the effective temperature of the element.

The secondary stars in binary MSPs and X-ray binaries are typically 
late-type stars, later than F, which contain metals absorption 
lines such as magnesium, calcium and iron as well as neutral hydrogen. 
The 
strongest absorption metal lines in the blue part of the optical spectrum 
are the Ca\,\textsc{I}\,4226\AA, Fe\,\textsc{I}\,4383\AA\ and 
Mg\,\textsc{I}\,b triplet 5167,5172,5183\AA. The EW versus temperature 
relation we use in our model is determined from observed stars in the 04 
to K2 spectral type range, obtained from the VLT-UVES Paranal Observatory 
Project database \citep{Bagnulo03}.  In Fig.\ref{fig:EW} we show the EW 
versus $T_{\rm eff}$ relation for the H$\alpha$ absorption line as also for 
comparison the Mg\,\textsc{I} triplet. 
Given that we do not really understand irradiation, the EW relation 
combined with $F_{\rm AV}$ allows the model 
to fit the observed line strength distribution 
\citep{Phillips99}.

\subsection{Additional sources of light}
\label{sec:extra}

Some of the optical light curves of binary MSPs are asymmetric with 
unequal maxima \citep{Stappers01, Schroeder14}. Various models have been 
proposed to explain this, such as off-centre heating from an intrabinary 
shock \citep{Romani16}, a hot spot \citep{Tang14} or solar-type star spots 
\citep{Staden16}.

The collision between the pulsar wind and the mass outflow from the 
secondary star can produce an intrabinary shock. Therefore the high energy 
radiation can be mediated by the intrabinary shock producing off-centre 
heating \citep{Li14,Deneva16}. Indeed the changes in colour and  spectral type 
across the face of the secondary star in \target\ does not match heating patterns,  
suggesting that the heating is not direct. 
X-ray observations of PSR\,J2215+5135 show an X-ray 
minimum near orbital phase zero (inferior conjunction of the secondary star) 
which has been interpreted as due to variable 
obscuration from an intrabinary shock around the secondary star 
\citep{Gentile14, Romani15}. In PSR\,J2215+5135, an intrabinary shock is suggested to 
explain the significant phase shift of the optical maximum with 
respect to the radio-pulse ephemeris, as well as the asymmetric optical 
light curves \citep{Romani16}.

In some MSPs, the transition from an accretion-powered to a 
rotation-powered pulsar has been observed, with the disappearance of 
features associated with an accretion disk, for example PSR\,J1023+0038 
\citep{Wang09}. However, as accretion onto the pulsar ceases, it is 
possible that a quiescent disk could remain between the companion and the 
light cylinder of the pulsar \citep{Eksi05}, which would contribute to the 
overall spectral energy distribution. In PSR\,J2215+5135, the difference 
between the observed colour temperature and the model temperature has been 
interpreted as an additional source of light from a hot accretion disk 
\citep{Schroeder14}. A large hot star spot could also produce an 
asymmetric temperature distribution.  It is also possible that the 
intrinsic magnetic field associated with the secondary star could channel 
the pulsar wind and cause enhanced local heating and an apparent hot spot 
on the star \citep{Tang14}.
Dark star spots due to strong magnetic activity on the secondary stars 
could also be present on the surface of the star \citep{Staden16}. Indeed 
Roche tomography has revealed observational evidence of star spots in 
binaries \citep{Shahbaz14}.
Therefore, to account for the possible sources of extra light we include an 
additional flux components in each wave-band and to simulate the effects of a 
dark star spot or off-centre heating (hot spot) which produces an 
asymmetric temperature distribution on the star, we add a Gaussian 
temperature function to the elements of area on the star. The 
normalization is positive or negative for a hot or dark sport, 
respectively. The Gaussian position and width is constrained in latitude 
and is extended uniformly in longitude across the star, roughly simulating 
a spot.

\begin{table}
\centering
\caption{Best-fit binary parameter Values for \target, based on the MCMC
\textsc{xrbcurve} analysis.}
\label{table:mcmc}
\begin{tabular}{lcc} 
\hline
Fitted Parameters        & Dark spot                   & Hot spot  \\
\hline
 $f$                     &   0.80\,$^{+0.03}_{-0.04}$  & 0.76\,$^{+0.03}_{-0.02}$    \\
 $q$                     &   0.18\,$^{+0.04}_{-0.04}$  & 0.28\,$^{+0.08}_{-0.09}$   \\
 $\cos\,i$  ($^{\circ}$) &   0.43\,$^{+0.06}_{-0.15}$  & 0.36$^{+0.06}_{-0.08}$    \\
 $K_{\rm 2}$  (\kms)     &  217.7\,$^{+7.4}_{-5.8}$    & 216.5$^{+5.8}_{-5.7}$   \\
 $\log F_{\rm X}$  (\ergs)    & -10.06\,$^{+0.09}_{-0.14}$   & --  \\
 $D_{\rm pc}$ (pc)       &    810\,$^{+85}_{-109}$     & 919\,$^{+109}_{-158}$      \\
 $T_{\rm 2}$  (K)        &   6902\,$^{+142}_{-223}$      & 6416$^{+209}_{-192}$        \\
 $T_{\rm spot}$ (K)      &    332 & 537  \\
 $A_{\rm spot}$ (\%)     &    2  & 3  \\ 
 $E_{\rm g'}$, $E_{\rm r'}$ and $E_{\rm i'}$ (\%) & 13, 11, 8 & 31, 25, 23 \\
% $E(B-V)$                &  0.36\,$^{+0.04}_{-0.05}$     &  0.25X\,$^{+0.X}_{-0.X}$  \\
 $\chi^2_{\rm min}$/DOF  &    183/122                &   193/123 \\
\hline      
Derived Parameters & \\
\hline      
$M_{\rm 1}$  ($M_{\sun}$)   & 1.72\,$^{+0.26}_{-0.29}$   & 1.85\,$^{+0.32}_{-0.26}$  \\
$M_{\rm 2}$  ($M_{\sun}$)   & 0.31\,$^{+0.10}_{-0.09}$   & 0.50\,$^{+0.22}_{-0.19}$  \\
$R_{\rm eq}$ ($R_{\sun}$)   & 1.18\,$^{+0.06}_{-0.08}$   & 1.16\,$^{+0.07}_{-0.08}$ \\
$v_{\rm rot}\,\rm sin\,{\it i}$ (\kms) 
                            &  62.5\,$^{+3.3}_{-3.1}$    & 63.1\,$^{+3.9}_{-3.9}$   \\
$\log  L_{\rm X}$ (\erg) & 33.8\,$^{+0.2}_{-0.2}$        & - \\
$\nu$; $F_\nu \propto \nu^\alpha$ & 1.1$\pm 0.3$ & 0.8$\pm 0.2 $ \\  
\hline
\end{tabular}
\end{table}

\section{\textsc{xrbcurve} fitting}
\label{sec:fitting}

Here we simultaneously fit the $g',r',i'$-band 
light curves and the absorption line radial velocity curve of \target\ 
presented in \citet{Linares16} with \textsc{xrbcurve} to determine the 
binary masses. The individual $g'$ (380 data points), $r'$ (629 data points), $i'$ (382 data points) band data 
points, were phase folded according to the orbital ephemeris and averaged 
into 39, 32 and 39 orbital phase bins, respectively. Similarly, the 131 
radial velocity points were averaged into 30 phase bins.

The optical light curves of \target\ clearly show unequal maxima. The fact 
that one sees a increase or decrease in light at phase 0.25 or 0.75, 
respectively, suggests that the extra source of light arises from the 
surface of the secondary star in terms of a dark star spot or a hot spot. 
To model these light curves with \textsc{xrbcurve} we assume either a hot 
or dark spot model. To reduce the number of free parameters in the model, 
we fix various model parameters. The value of the gravity-darkening 
exponent depends on the mean temperature and gravity of the secondary star 
\citep{Claret11}. The observed F6 late-type spectral type for the 
secondary star allows us to fix the gravity-darkening exponent to 0.072 
\citep{Lucy67, Claret11}. 
From the observed hydrogen column density we 
find the reddening to be $E(B-V)$=0.251$\pm$0.054 \citep{Linares16}.

The binary model parameters are $f$, $q$, $\cos\,i$, $T_{\rm 2}$, $\log 
F_{\rm X}$, $D_{\rm pc}$, $K_{\rm 2}$, $F_{\rm AV}$ and $E(B-V)$.
There is also the additional source 
of light in each filter $E_{\rm g'}$, $E_{\rm r'}$ and $E_{\rm i'}$, the 
light curve phase shift $\delta_{\rm g'}$, $\delta_{\rm r'}$ and 
$\delta_{\rm i'}$, the radial velocity phase shift and systemic velocity 
$\delta_{\rm AV}$, $\gamma_{\rm AV}$, and finally the Gaussian spot 
parameters, position $C_{\rm spot}$, width $W_{\rm spot}$, normalization 
$T_{\rm spot}$ and extent $E_{\rm spot}$. 

Given $q$, $\cos\,i$ and $K_{\rm 
2}$ we calculate $M_{\rm 1}$ and $M_{\rm 2}$. Given $q$ and $f$ we 
determine $R_{\rm eq}$ and which when combined with $q$ and $K_{\rm 2}$ 
gives \vsini\ (see Section\,\ref{sec:evr}). However, note that the observed 
value for \vsini\ depends on orbital phase, because it reflects the size 
of the tidally distorted star.
The value for $R_{\rm eq}$ we determine represents the mean radius of the 
star observed and so the predicted value for \vsini\  is a 
mean value.

To optimize the fitting procedure to 
first fit photometric light curve and the absorption-line radial 
velocity curve with \textsc{xrbcurve} assuming a dark spot 
 on the 
secondary star. We use the differential evolution algorithm described in 
\citet{Shahbaz03a} with a crossover probability of 0.9, 
a  mutation scaling factor of 0.9, and a 
 maximum number of generations of 2000, 
which corresponds to $\sim$ 100,000 calculations of the 
model. Given that there are a number of different types of data with 
different number of data points, to optimize the fitting procedure we 
assigned relative weights to the different data sets, determined after an 
initial fit. 
We rescaled the error bars on 
each data set so 
that the total reduced $\chi^2$ of the fit was $\sim 1$ for each data set 
separately.  

The simultaneous light and radial velocity curve fitting is a 
multi-dimensional nonlinear optimization problem. In order to obtain 
a robust error analysis 
we used Markov chain Monte Carlo (MCMC) method convolved with a differential 
evolution fitting algorithm; \textsc{d}iffe\textsc{r}ential 
\textsc{e}volution \textsc{a}daptive \textsc{m}etropolis (\textsc{dream}) 
\citep[see][and references within]{Vrugt2016}, which works well in problems with 
a high number of  dimensions. 
\textsc{dream} simultaneously runs multiple 
different chains uses differential evolution \citep{Storn95} as 
the genetic algorithm for population evolution with a Metropolis selection 
rule to decide whether parents are replaced by 
candidate points not. 
In order to ensure that the MCMC has converged 
we also performed the Gelman-Ruben test, 
which analyzes the difference between multiple 
Markov chains and we also visually inspected 
the trace of the parameters. 
The convergence is assessed by comparing the estimated 
between-chains and within-chain variances for each model parameter 
\citep{Gelman92}.
We use a Bayesian framework to make statistical inferences on our binary 
model parameters. 
Our fitting makes use of flat prior probability 
distributions for all the model parameters, except $E(B-V)$, where we use 
a Gaussian prior
[see \citet{Gregory05} for an 
introduction to Bayesian analysis]. 
We use 20 individual chains to explore 
the parameter space and followed each for 40,000 iterations.  We reject 
the first 500 iterations ("burn-in") and only include every 10th point 
("thinning"), based on the observed auto correlation lengths for the 
individual parameters.

%%%%%%%%%%%%%%%%%%%%%%%%%%%%%%%%%%%%%%%%%%%%%%%%%%%%%%%%%%%%%%%%%%%%%%%
\begin{figure*}
\includegraphics[width=\linewidth]{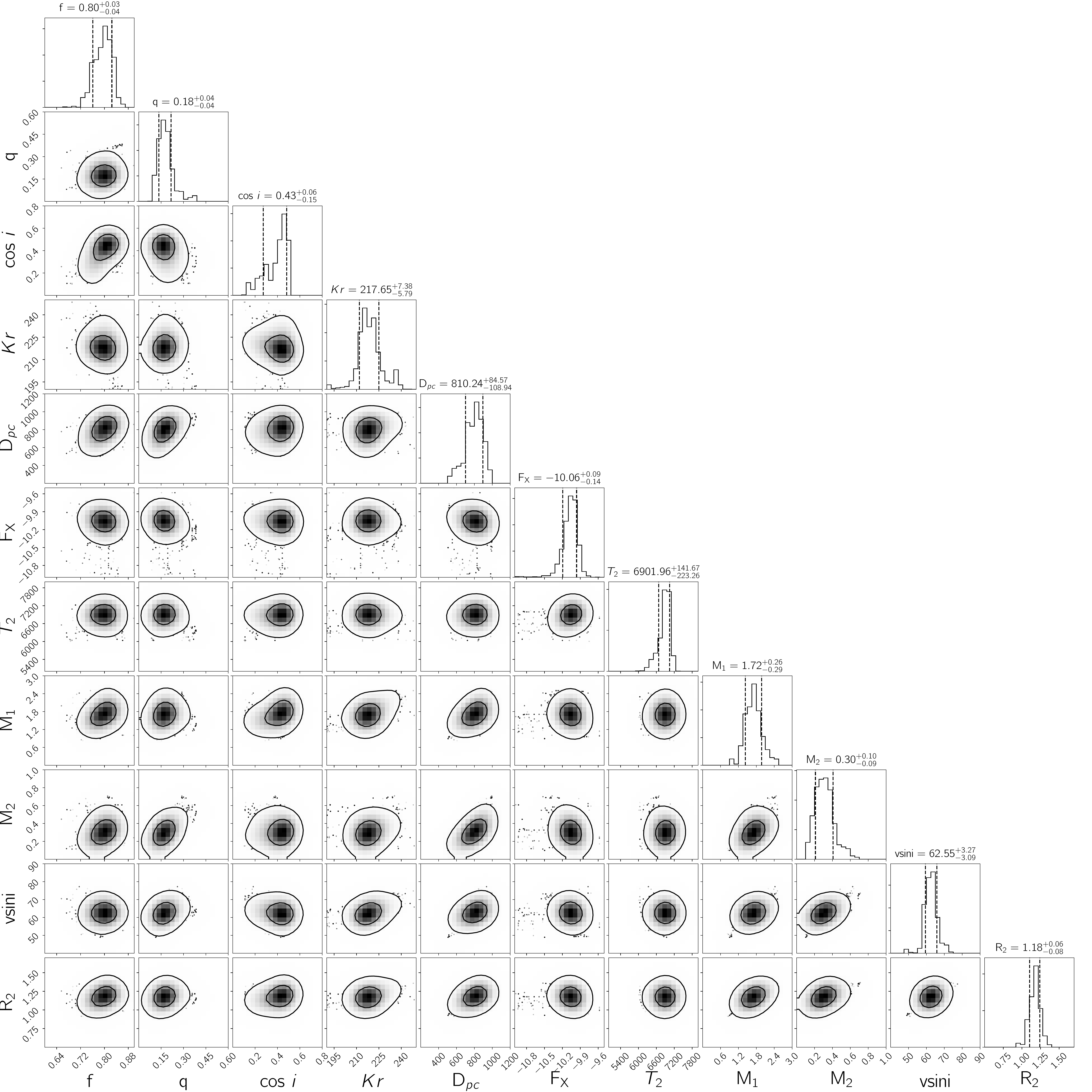}
\caption{One- and two-dimensional distributions of the model parameter 
values resulting from the MCMC fitting for the model with a dark spot. 
The mean and standard deviation of each 
parameter are given. 
The vertical dashed lines in the 1-D histogram plots show the 
1-$\sigma$ limits on the derived parameter. 1 and 2-$\sigma$ 
contours are shown in the 2-D plots. 
The mass of the pulsar and secondary star, the 
secondary star's volume equivalent radius and its projected rotational 
velocity are inferred values from radial velocity semi-amplitude, binary 
mass ratio, Roche lobe filling factor and inclination angle. }
\label{fig:mcmc_dark}
\end{figure*}
%%%%%%%%%%%%%%%%%%%%%%%%%%%%%%%%%%%%%%%%%%%%%%%%%%%%%%%%%%%%%%%%%%%%%%%

%%%%%%%%%%%%%%%%%%%%%%%%%%%%%%%%%%%%%%%%%%%%%%%%%%%%%%%%%%%%%%%%%%%%%%%
\begin{figure*}
\includegraphics[width=\linewidth]{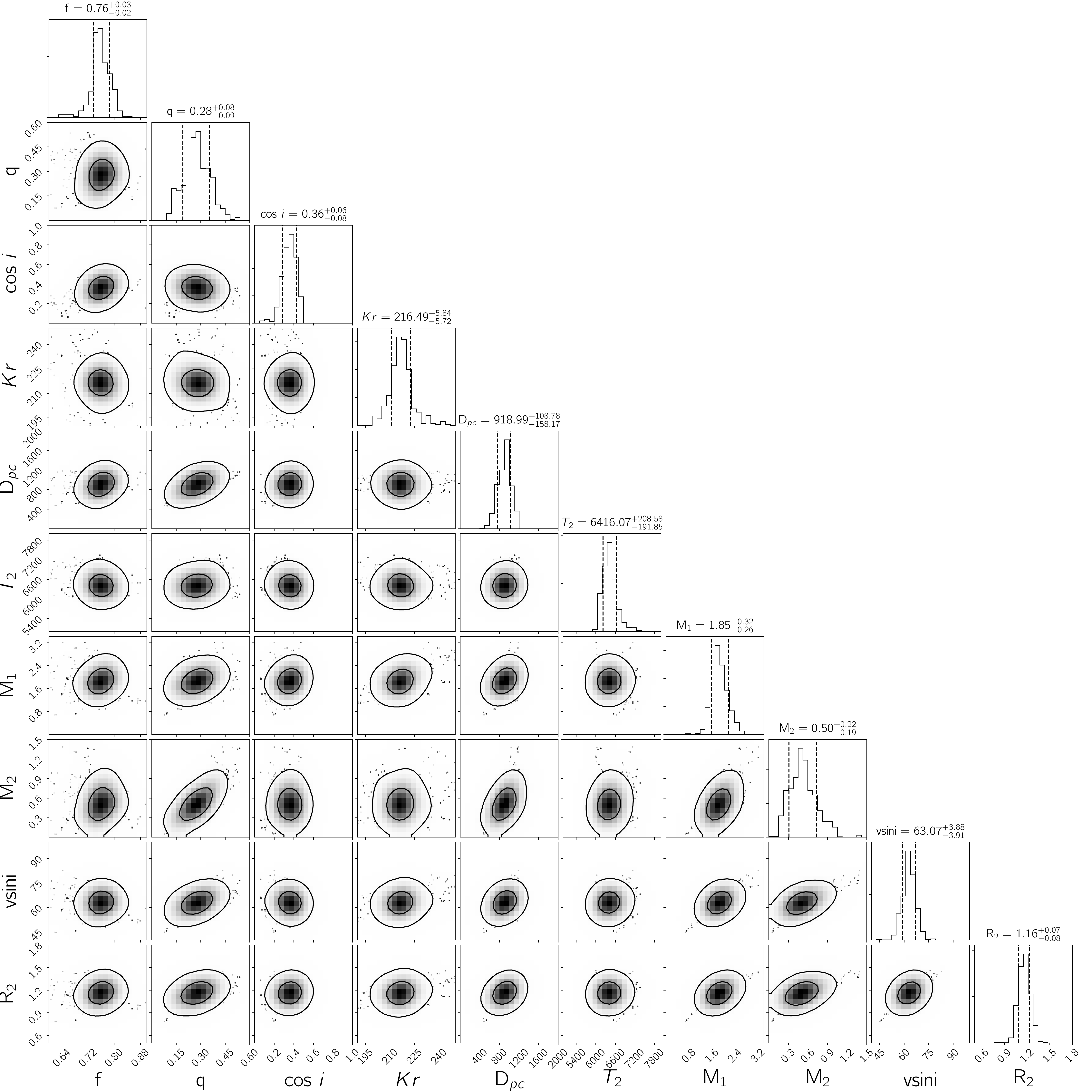}
\caption{ Same as Fig.\,\ref{fig:mcmc_dark} but for the hot spot model. }
\label{fig:mcmc_hot}
\end{figure*}
%%%%%%%%%%%%%%%%%%%%%%%%%%%%%%%%%%%%%%%%%%%%%%%%%%%%%%%%%%%%%%%%%%%%%%%

\section{Results}
\label{sec:results}

% http://www.six-sigma-material.com/Normal-Distribution.html

We simultaneously fit $g'$, $r'$ and $i'$-band light curves and an 
absorption line radial velocity curve of \target\ with \textsc{xrbcurve} 
with a constant source of light in each band to account for a possible 
disk and/or intrabinary shock and two model scenarios (a) a dark star spot or 
(b) a hot spot due to off-centre heating  to account for the unequal maxima.
Our preliminary fits show that the dark or hot spot model $F_{\rm AV}$ is not 
constrained, which means that the model requires all the inner face of the 
star contributes to the radial velocity curve, so we fix it to a large value 
accordingly.
Furthermore, we find that the  phase shifts for each filter are the same, 
so in our final fits we set  
$\delta_{\rm i'}$=$\delta_{\rm r'}$=$\delta_{\rm g'}$.

Also, for the hot spot model
we find that no heating is required, so we fix $F_{\rm X}$ to zero. 
The dark and hot spot model have 18 and 17 free model parameters, respectively.
For our final fits, the dark and hot spot model best fits have a $\chi^2$ 
of 182 and 193, with 122 and 123 degrees of freedom,  respectively. 
Using the F-test to test the null hypothesis that the $\chi^2$ of two 
models are the same, we find that we can reject the null hypothesis at the 
1.1 per cent significance level, 
Hence, statistically the dark spot 
model is only better than the 
hot spot model at the 98.9 per cent  ($\sim2.29\sigma$) confidence level.

The best fit models for the hot and dark spot are 
shown in Figures\,\ref{fig:hot} and \ref{fig:dark}, respectively. The plots 
shows the best fit to the light and radial velocity curves as well as 
the light curve of the extra source of light in each band. We also show 
the observed maps of the continuum flux and H$\alpha $absorption line strength 
on the secondary star at different orbital phases. In 
Table\,\ref{table:mcmc} we give the mean and 1-$\sigma$ limits of the 
posterior probability distribution functions on each parameter for both 
models. The result of the MCMC for the hot and dark spot model are shown 
in Figs.\,\ref{fig:mcmc_hot} and \ref{fig:mcmc_dark}, respectively.  As one can see, most 
of the two-dimensional  probability distribution functions are relatively 
well determined. 
Changing the gravity darkening coefficient only changes the $\chi^2$ by  
3, resulting in a change in $M_1$ of 10 per cent.

\begin{figure}
\centering
\includegraphics[width=\columnwidth, angle=-90, scale=0.5]{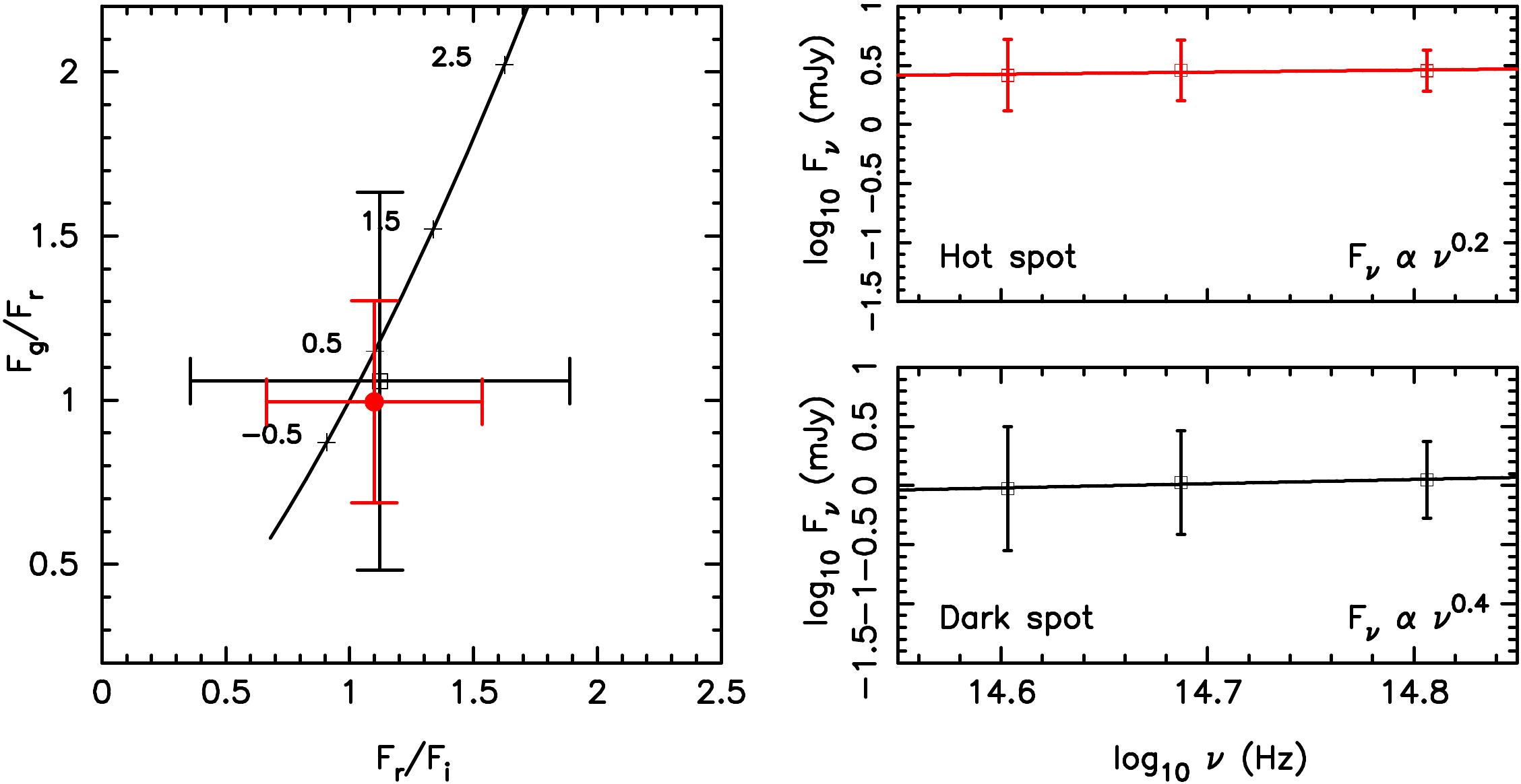}
\caption{The spectrum of the extra flux contribution determined from the 
hot (red solid circles) and dark (black open squares) spot model. The left 
plot shows the colour-colour diagram where the solid line shows flux 
ratios for different power-law spectrum energy distributions of the form 
$F_\nu \propto \nu^\alpha$. The right plots show the spectral energy 
distribution with the best fit power-law. }
\label{fig:extra} 
\end{figure}

The fits with the dark and hot 
spot model give similar values for $f$, $i$ and $K_{\rm 2}$ (see 
Table\,\ref{table:mcmc}). 
However, the value for $q$ and $T_{\rm 2}$ are different.  The dark spot 
model best fit suggests a system at an inclination angle of 65$^\circ$ and mass 
ratio $q$=0.18 with some effects of X-ray heating; 24 percent 
of the secondary star has a change in temperature more than 10 degrees 
larger than the non-irradiated level. The dark spot has a minumum 
temperature of $T_{\rm spot}$=332\,K and covers 2
percent of the star; see Fig.\,\ref{fig:dark}). In contrast the best fit with a 
hot spot suggest a system at slightly higher inclination angle of 69$^\circ$ and 
a less extreme mass ratio $q$=0.28 with no heating. 
The hot spot has a temperature of $T_{\rm spot}$=538\,K 
and covers 3 percent of the star.
For the dark spot model we find that the extra source of light contributes 
13, 11 and 8  per cent to the observed flux in the $g'$, $r'$ and $i'$ 
bands, respectively. Similarly for the hot spot model we find that the 
extra source of light contributes 31, 25 and 23  per cent to the observed 
flux in the $g'$, $r'$ and $i'$ bands, respectively. In 
Fig.\,\ref{fig:extra} we show the spectrum of the extra flux component for 
the dark and hot spot model, which can be represented with a spectral index 
($F_\nu\propto \nu^\alpha$) of 1.1$\pm 0.3$ and 0.8$\pm 0.2$, respectively.
     
In principle the shape of the ellipsoidal light curve can provide 
important clues to the binary inclination angle, because the peak to peak 
amplitude and difference between the two minima depend primarily on the 
binary inclination angle and mass ratio. Irradiation not only fills the 
minimum at phase 0.5, but it also a produces a larger amplitude 
modulation. 
However, for a Roche lobe under-filling star, this dependence 
with mass ratio disappears, because the star is not as tidally distorted 
compared to a fully Roche lobe filling star. For our values of $f$ the 
dependence of the light curve amplitude with mass ratio still exists.
\citep{Shahbaz98}. For a similar inclination angle, a system with a less extreme
mass ratio and no heating will produce a similar light curve to a 
system with a more extreme mass ratio and  with  heating. Our hot and dark 
spot model fits suggest either a non-irradiated system at an extreme mass ratio  
or weakly irradiated system at a less extreme mass ratio, respectively. 

The lack of heating of the secondary star 
required by the hot spot model by the pulsar wind and/or radiation is not 
surprising, given the wide orbit. The only other MSP which also has a wide 
orbit with a similar lack of irradiation is PSR\,J1740--5340 
\citep{Orosz03}. 
Both the hot and dark spot model give similar values for 
$f$, $K_{\rm 2}$, $i$ and \vsini, but different values for $q$ and  
$T_{\rm 2}$. $T_{\rm 2}$ determined from the hot spot model 
is consistent with the observed 
spectral type F6$\pm$2 \citep{Linares16}, corresponding to a temperature 
range of 6170--6640\,K \citep{Pecaut13}.

%%%%%%%%%%%%%%%%%%%%%%%%%%%%%%%%%%%%%%%%%%%%%%%%%%%%%%%%%%%%%%%%%%%%%%%
%%%%%%%%%%%%%%%%%%%%%%%%%%%%%%%%%%%%%%%%%%%%%%%%%%%%%%%%%%%%%%%%%%%%%%%

\section{Discussion}

\subsection{The extra light source}

\citet{Zelati14} determined the spectral energy distribution of 
PSR\,J1023+0038 from near-IR to X-rays, when the system was in an 
accretion-powered phase. They found that the spectral energy distribution 
is well modelled by contributions from the secondary star, the accretion 
disc and an intrabinary shock. The neutron star spin-down luminosity 
irradiates the accretion disc and the secondary star, accounting for the 
UV and optical emission. X-rays and gamma-rays are produced in an 
intrabinary shock and the shock emission is powered by the neutron star 
spin-down luminosity which extends to lower energies with a photon index 
of $\Gamma \sim$1.5 which corresponds to a spectral index of $\alpha 
\sim$2.5 ($\alpha = 1+\Gamma$; $EN(E) \propto f_\nu; N(E) \propto 
E^\Gamma; F_\nu \propto \nu^\alpha$, where $N(E)$ is the X-ray (0.5--10 keV) 
photon spectrum 
and $f_\nu$ is the source spectrum). Indeed, the photon index measured for 
redbacks in the accretion or pulsar state is in the range 0.9--1.8 
\citep{Linares14} and \target\ is no exception with a photon index of 
$\Gamma \sim 1.3$, corresponding to a spectral index of $\alpha \sim 2.3$.

When determining \vsini\ \citet{Linares16} found that their best match 
template star required a non-stellar light veiling of 10--30 per cent in 
the $\sim g'$-band. For the dark spot model we find that the extra source 
of light contributes 13, 11 and 8 per cent to the observed flux in the $g'$, 
$r'$ and $i'$ bands, respectively. Similarly for the hot spot model we 
find that the extra source of light contributes 31, 25 and 23 per cent to 
the observed flux in the $g'$, $r'$ and $i'$ bands, respectively. Both models 
agree well with the veiling measured from the spectra \citep{Linares16}.
The extra light source contribution required to fit our light curves can be 
represented with a spectral index; $F_\nu\propto \nu^\alpha$. We find a spectral 
index of 1.1 and 0.8, for the dark and hot spot model, respectively, which does not
 agree with the X-ray spectral index, suggesting that the extra optical emission 
may not be produced by an intrabinary shock.

\subsection{Energy flux}

\citet{Li16} and \citet{Linares16} have determined the spectral energy 
distribution of \target\ which shows the optical band dominated by the 
secondary star, the gamma-rays from the MSP and the shock between the MSP 
and secondary star's wind (intrabinary shock), which most likely powers 
the X-ray emission. The unabsorbed 0.1--100\,GeV 
energy flux is 1.71$\times$10$^{-11}$\ergs, whereas the unabsorbed 
0.5--10\,keV luminosity is 1.8$\times$10$^{-12}$\ergs 
\citep{Acero15, Linares16}. We can compare these values with the bolometric 
flux  determined from our dark spot model; no X-ray heating is 
required in the hot spot model.
For the  dark spot model we find a flux of  6.6$\times$10$^{-11}$\ergs which 
is a factor of $\sim$4 more than what is observed in the $\gamma$-rays. 
The extra source of heating arise from the pulsar which 
emits prompt particles to heat the secondary star \citep{Romani16}

\subsection{The secondary stars rotational broadening}
\label{sec:vsini}

Synchronization via tidal forces will suppress differential rotation 
and make the angular velocity of the secondary star constant.
However, since the star is distorted, the linear rotational 
velocity will vary with longitude around the star and so \vsini\ will vary 
across the orbit. The variations show two maxima and two minima per 
orbital cycle and have an amplitude of $\sim$10\kms, which depends on $q$ 
and $i$ \citep{Shahbaz98}. The procedure normally used to measure the 
secondary star's rotational broadening is to compare it with the spectrum 
of a slowly rotating template star convolved with a limb-darkened standard 
rotation profile \citep[see][and references within]{Gray05, Collins95}. 
The width of the 
limb-darkened standard rotation 
profile (adopting the continuum value for the limb-darkening coefficient) is varied until it matches the width of the target spectrum 
\citep[see, for example][]{Marsh94}. 
However, because of the 
assumed spherical shape of the rotation profile. 
there are assumptions inherent in this method.
The secondary stars in 
binary MSPs substantially fill their Roche lobes, and so they will have distorted 
line profiles which depend on the exact Roche binary geometry \citep{Shahbaz98}. 
Also, because both temperature and gravity vary over the star's photosphere 
due to the Roche lobe shape, the spectrum of the secondary 
cannot be described by a single-star spectrum. 
Finally, we use the 
continuum value for the limb-darkening coefficient.
Firstly, because the line flux arises from higher regions in the 
atmosphere than the continuum flux, absorption lines have core 
limb-darkening coefficients much less than the continuum value \citep{Collins95}. 
Hence using the standard rotation profile with zero and the continuum value 
for the line 
limb-darkening coefficient gives a value for $q$ that brackets the value 
found using the full geometrical treatment \citep{Shahbaz03b}. Using the 
extreme cases for the limb-darkening coefficient of zero or the continuum 
value introduces a systematic uncertainty of about 14 per cent in the 
determination \vsini\ \citep{Welsh95}.

\citet{Linares16} used the standard method described above to estimate the 
projected rotational velocity of the secondary star, which they find to be 
\vsini=73.2$\pm$1.6\kms from spectra around orbital phase 0.2. They then 
use this \vsini\ with their value for $K_{\rm 2}$, determined from a 
sinusoidal fit to their radial velocity curve and obtain $q$=0.26 (see 
equation \,\ref{eq:1}), assuming a Roche lobe filling, tidally locked and 
spherically symmetric companion star (see Section\,\ref{sec:evr}). 
However, the variation of \vsini\ with orbital phase (10--15 per cent) 
introduces coupled with the uncertainties in the line limb-darkening 
coefficient ($\sim$14 per cent) and the fact that the star does not fill 
its Roche lobe, contributes significantly to the accuracy to which one can 
determine $q$ using this method.

\subsection{Favoured model}

\target\ is thought to be a redback based on its optical, 
X-ray and $\gamma$-ray properties, 
which are similar to other redbacks \citep[see Table 1 in][]{Li16}. Most redbacks 
have secondary stars with minimum secondary star masses of $<$0.8\msun  
\citep{Roberts13, Strader14} and so the secondary star masses determined from  
the dark and hot spot model are consistent with the system being a redback. 
{\ The reduced $\chi^2$ of the best-fit for the dark and hot spot models 
are significantly different, with the dark spot model being only statistically 
better at the 98.9 (=2.29$\sigma$) confidence level (see Section\,\ref{sec:results}), 
The value determined for \vsini\ for both models agrees (within $\sim$3-$\sigma$) with
the observations (73.2$\pm$1.6\kms\, \citealt{Linares16}), however, the hot
spot model with $T_{\rm 2}$=6416\,K more constent with observations. 
(6170--6640\,K \citet{Linares16}). We therefore, favour the
hot spot model which gives  a neutron star mass of 
$M_{\rm 1}$=1.85\,$^{+0.32}_{-0.26}$ \msun\ and a secondary star mass of 
$M_{\rm 2}$=0.50\,$^{+0.22}_{-0.19}$ \msun.

}

\subsection{Binary masses}

\citet{Li16} obtained $R$ and $g'$ band light curves and two absorption  line
radial velocity points. They used the \textsc{ELC} code  \citep{Orosz00} to fit
the data, assuming $T_{\rm 2}$=5750\,K, $M_{\rm 2}  \sim$0.4\,$M_{\sun}$ with no
irradiation effects. They performed a grid  search in $i$=60--90\,degrees and
found $\beta_V$=0.70 to 0.64 (where  $\beta_V$ is the ratio of the equivalent volume
radius of the companion star  to the Roche lobe radius) corresponding to $M_{\rm
1}$=1.5 to  2.2\,$M_{\sun}$. Note that their definition for the Roche lobe
filling  factor is different to that in our model. Their estimated range for 
$\beta_V$ corresponds to $f$=0.50 to 0.54 (determined using \textsc{xrbcurve}), 
which is very different to the value we obtain (see  Table\,\ref{table:mcmc}).
However, as noted by the authors, given their  poor fits and large systematics,
the binary parameters they obtain are  merely indicative. In contrast we simultaneously
fit the absorption line radial  velocity and $g'$, $r'$, $i'$ light curve, which
allow us to constrain various model parameters such as the filling factor of the
companion and its surface temperature. It is evident that this redback companion
does not fill its Roche lobe, thus confirming the previous finding  by
\citet{Breton13} that redbacks and black widows may have smaller radii than
originally expected (due to the presence of radio eclipses in some systems).

The presence of temperature asymmetry also indicates that a hot or cold spot is
present on the companion's surface. The physical reason for this remains
unclear  (see Section\,\ref{sec:extra} for possible models), but long-term
monitoring could provide further insights to shed light on this mystery. A cold
or a hot spot due to an atmospheric phenomenon might move on a time scale
related to the magnetic activity on the star for instance.  Star spots are 
believed to be common features on many late-type stars
and typically vary on timescales of months to a few years \cite{Bouvier89}. 
Star spots have been observed in the surface map of the secondary star in the
LMXB Cen\,X--4  \citep{Shahbaz14}. Indeed, in cataclysmic variables, studies of
the surface maps of the secondary stars show star spots moving on timescale of
few days  \citep{Hill14}.

There are about 32 reliable neutron star mass measurements in X-ray binaries, 
double neutron star systems and millisecond pulsars, most of which are 
are at least partly based on the radio timing techniques \citep{Ozel16},
Studies statistically distinguish  between different types of neutron stars and 
 between those believed to be close to 
their birth masses, with a mean mass of 1.35$\pm$0.05\msun\  which reflects a highly tuned 
formation channel \citep{Ozel12, Kiziltan13}, and the ones that have undergone  
long-term accretion episodes.
Most recycled pulsars are accompanied by mid-F to late-K type low-mass 
$<$0.5\msun\ companions but for these systems to form in the age of the 
Universe, the companion must originally have been a star of $\sim$1\msun. 
Thus, the companion star must have lost a large fraction of 
its mass, accreted onto the neutron star, leading to a increase in the neutron 
star's mass. The most precise massive neutron star measurements have been 
obtained through the measurement of the Shapiro delay, 1.94$\pm$0.04\msun\ and 
1.667$\pm$0.021\msun\ in PSR\,J1614-2230 and PSR\,J1903+0327, respectively 
\citep{Demorest10, Freire11}. These masses has allowed us to place fundamental constraints 
on the equation of state of nuclear matter at high densities, excluding 
many of the soft equations of state \citep{Lattimer04}. Dynamical photometric 
and spectroscopic studies of binary MSPs have largely revealed that 
neutron star masses in these systems are generally heavier than the 
1.4\,$M_{\odot}$ canonical value, 
with some particularly heavy ones in PSR\,B1957+20 ($M_{\rm 1}$=2.4\msun, 
\citealt{Kerkwijk11}) and PSR\,J1816+4510 ($M_{\rm 1}\geq$1.84\msun, 
\citealt{Kaplan13}). 
Our mass measurement for the neutron star in \target\ using  the 
hot spot model suggests a massive 1.85\,$^{+0.32}_{-0.26}$\msun\ neutron star.

\section*{ACKNOWLEDGEMENTS}

This research has been supported by the Spanish Ministry of Economy and 
Competitiveness (MINECO) under the grant AYA2013-42627.
M.L. is supported by EU's Horizon 2020 programme through a 
Marie Sklodowska-Curie Fellowship (grant nr. 702638). R.P.B. received
funding  from  the  European  Union  Seventh  Framework
Programme under grant agreement PIIF-GA-2012-332393.
This paper makes use of the IAC's Supercomputing facility
\textsc{condor}.

 \newcommand{\noop}[1]{}


\begin{thebibliography}{}
\makeatletter
\relax
\def\mn@urlcharsother{\let\do\@makeother \do\$\do\&\do\#\do\^\do\_\do\%\do\~}
\def\mn@doi{\begingroup\mn@urlcharsother \@ifnextchar [ {\mn@doi@}
  {\mn@doi@[]}}
\def\mn@doi@[#1]#2{\def\@tempa{#1}\ifx\@tempa\@empty \href
  {http://dx.doi.org/#2} {doi:#2}\else \href {http://dx.doi.org/#2} {#1}\fi
  \endgroup}
\def\mn@eprint#1#2{\mn@eprint@#1:#2::\@nil}
\def\mn@eprint@arXiv#1{\href {http://arxiv.org/abs/#1} {{\tt arXiv:#1}}}
\def\mn@eprint@dblp#1{\href {http://dblp.uni-trier.de/rec/bibtex/#1.xml}
  {dblp:#1}}
\def\mn@eprint@#1:#2:#3:#4\@nil{\def\@tempa {#1}\def\@tempb {#2}\def\@tempc
  {#3}\ifx \@tempc \@empty \let \@tempc \@tempb \let \@tempb \@tempa \fi \ifx
  \@tempb \@empty \def\@tempb {arXiv}\fi \@ifundefined
  {mn@eprint@\@tempb}{\@tempb:\@tempc}{\expandafter \expandafter \csname
  mn@eprint@\@tempb\endcsname \expandafter{\@tempc}}}

\bibitem[\protect\citeauthoryear{{Abdo}}{{Abdo}}{2009}]{Abdo09}
{Abdo} A.~A. e.~a.,  2009, \mn@doi [Science] {10.1126/science.1176113}, \href
  {http://adsabs.harvard.edu/abs/2009Sci...325..848A} {325, 848}

\bibitem[\protect\citeauthoryear{{Acero}}{{Acero}}{2015}]{Acero15}
{Acero} F. e.~a.,  2015, \mn@doi [\apjs] {10.1088/0067-0049/218/2/23}, \href
  {http://adsabs.harvard.edu/abs/2015ApJS..218...23A} {218, 23}

\bibitem[\protect\citeauthoryear{{Alpar}, {Cheng}, {Ruderman}  \&
  {Shaham}}{{Alpar} et~al.}{1982}]{Alpar82}
{Alpar} M.~A.,  {Cheng} A.~F.,  {Ruderman} M.~A.,   {Shaham} J.,  1982, \mn@doi
  [\nat] {10.1038/300728a0}, \href
  {http://adsabs.harvard.edu/abs/1982Natur.300..728A} {300, 728}

\bibitem[\protect\citeauthoryear{{Archibald} et~al.,}{{Archibald}
  et~al.}{2009}]{Archibald09}
{Archibald} A.~M.,  et~al., 2009, \mn@doi [Science] {10.1126/science.1172740},
  \href {http://adsabs.harvard.edu/abs/2009Sci...324.1411A} {324, 1411}

\bibitem[\protect\citeauthoryear{{Bagnulo}, {Jehin}, {Ledoux}, {Cabanac},
  {Melo}, {Gilmozzi}  \& {ESO Paranal Science Operations Team}}{{Bagnulo}
  et~al.}{2003}]{Bagnulo03}
{Bagnulo} S.,  {Jehin} E.,  {Ledoux} C.,  {Cabanac} R.,  {Melo} C.,  {Gilmozzi}
  R.,   {ESO Paranal Science Operations Team} 2003, The Messenger, \href
  {http://adsabs.harvard.edu/abs/2003Msngr.114...10B} {114, 10}

\bibitem[\protect\citeauthoryear{{Bassa} et~al.,}{{Bassa}
  et~al.}{2014}]{Bassa14}
{Bassa} C.~G.,  et~al., 2014, \mn@doi [\mnras] {10.1093/mnras/stu708}, \href
  {http://adsabs.harvard.edu/abs/2014MNRAS.441.1825B} {441, 1825}

\bibitem[\protect\citeauthoryear{{Benvenuto}, {De Vito}  \&
  {Horvath}}{{Benvenuto} et~al.}{2014}]{Benvenuto14}
{Benvenuto} O.~G.,  {De Vito} M.~A.,   {Horvath} J.~E.,  2014, \mn@doi [\apjl]
  {10.1088/2041-8205/786/1/L7}, \href
  {http://adsabs.harvard.edu/abs/2014ApJ...786L...7B} {786, L7}

\bibitem[\protect\citeauthoryear{{Bhattacharya} \& {van den
  Heuvel}}{{Bhattacharya} \& {van den Heuvel}}{1991}]{Bhattacharya91}
{Bhattacharya} D.,  {van den Heuvel} E.~P.~J.,  1991, \mn@doi [\physrep]
  {10.1016/0370-1573(91)90064-S}, \href
  {http://adsabs.harvard.edu/abs/1991PhR...203....1B} {203, 1}

\bibitem[\protect\citeauthoryear{{Bouvier} \& {Bertout}}{{Bouvier} \&
  {Bertout}}{1989}]{Bouvier89}
{Bouvier} J.,  {Bertout} C.,  1989, \aap, \href
  {http://adsabs.harvard.edu/abs/1989A%26A...211...99B} {211, 99}

\bibitem[\protect\citeauthoryear{{Breton} et~al.,}{{Breton}
  et~al.}{2013}]{Breton13}
{Breton} R.~P.,  et~al., 2013, \mn@doi [\apj] {10.1088/0004-637X/769/2/108},
  \href {http://adsabs.harvard.edu/abs/2013ApJ...769..108B} {769, 108}

\bibitem[\protect\citeauthoryear{{Chen}, {Chen}, {Tauris}  \& {Han}}{{Chen}
  et~al.}{2013}]{Chen13}
{Chen} H.-L.,  {Chen} X.,  {Tauris} T.~M.,   {Han} Z.,  2013, \mn@doi [\apj]
  {10.1088/0004-637X/775/1/27}, \href
  {http://adsabs.harvard.edu/abs/2013ApJ...775...27C} {775, 27}

\bibitem[\protect\citeauthoryear{{Claret} \& {Bloemen}}{{Claret} \&
  {Bloemen}}{2011}]{Claret11}
{Claret} A.,  {Bloemen} S.,  2011, \mn@doi [\aap]
  {10.1051/0004-6361/201116451}, \href
  {http://cdsads.u-strasbg.fr/abs/2011A%26A...529A..75C} {529, A75}

\bibitem[\protect\citeauthoryear{{Collins} \& {Truax}}{{Collins} \&
  {Truax}}{1995}]{Collins95}
{Collins} II G.~W.,  {Truax} R.~J.,  1995, \mn@doi [\apj] {10.1086/175225},
  \href {http://adsabs.harvard.edu/abs/1995ApJ...439..860C} {439, 860}

\bibitem[\protect\citeauthoryear{{Coti Zelati} et~al.,}{{Coti Zelati}
  et~al.}{2014}]{Zelati14}
{Coti Zelati} F.,  et~al., 2014, \mn@doi [\mnras] {10.1093/mnras/stu1552},
  \href {http://adsabs.harvard.edu/abs/2014MNRAS.444.1783C} {444, 1783}

\bibitem[\protect\citeauthoryear{{Crawford} et~al.,}{{Crawford}
  et~al.}{2013}]{Crawford13}
{Crawford} F.,  et~al., 2013, \mn@doi [\apj] {10.1088/0004-637X/776/1/20},
  \href {http://adsabs.harvard.edu/abs/2013ApJ...776...20C} {776, 20}

\bibitem[\protect\citeauthoryear{{Demorest}, {Pennucci}, {Ransom}, {Roberts}
  \& {Hessels}}{{Demorest} et~al.}{2010}]{Demorest10}
{Demorest} P.~B.,  {Pennucci} T.,  {Ransom} S.~M.,  {Roberts} M.~S.~E.,
  {Hessels} J.~W.~T.,  2010, \mn@doi [\nat] {10.1038/nature09466}, \href
  {http://adsabs.harvard.edu/abs/2010Natur.467.1081D} {467, 1081}

\bibitem[\protect\citeauthoryear{{Deneva} et~al.,}{{Deneva}
  et~al.}{2016}]{Deneva16}
{Deneva} J.~S.,  et~al., 2016, \mn@doi [\apj] {10.3847/0004-637X/823/2/105},
  \href {http://adsabs.harvard.edu/abs/2016ApJ...823..105D} {823, 105}

\bibitem[\protect\citeauthoryear{{Eggleton}}{{Eggleton}}{1983}]{Eggleton83}
{Eggleton} P.~P.,  1983, \mn@doi [\apj] {10.1086/160960}, \href
  {http://adsabs.harvard.edu/abs/1983ApJ...268..368E} {268, 368}

\bibitem[\protect\citeauthoryear{{Ek{\c s}{\.I}} \& {Alpar}}{{Ek{\c s}{\.I}} \&
  {Alpar}}{2005}]{Eksi05}
{Ek{\c s}{\.I}} K.~Y.,  {Alpar} M.~A.,  2005, \mn@doi [\apj] {10.1086/425959},
  \href {http://adsabs.harvard.edu/abs/2005ApJ...620..390E} {620, 390}

\bibitem[\protect\citeauthoryear{{Freire} et~al.,}{{Freire}
  et~al.}{2011}]{Freire11}
{Freire} P.~C.~C.,  et~al., 2011, \mn@doi [\mnras]
  {10.1111/j.1365-2966.2010.18109.x}, \href
  {http://adsabs.harvard.edu/abs/2011MNRAS.412.2763F} {412, 2763}

\bibitem[\protect\citeauthoryear{Gelman \& Rubin}{Gelman \&
  Rubin}{1992}]{Gelman92}
Gelman A.,  Rubin D.,  1992, Statistical Science, 7, 457

\bibitem[\protect\citeauthoryear{{Gentile} et~al.,}{{Gentile}
  et~al.}{2014}]{Gentile14}
{Gentile} P.~A.,  et~al., 2014, \mn@doi [\apj] {10.1088/0004-637X/783/2/69},
  \href {http://adsabs.harvard.edu/abs/2014ApJ...783...69G} {783, 69}

\bibitem[\protect\citeauthoryear{Gray}{Gray}{2005}]{Gray05}
Gray D.~F.,  2005, The Observation and Analysis of Stellar Photospheres:, 3
  edn.
Cambridge University Press, Cambridge, \mn@doi{10.1017/CBO9781316036570}

\bibitem[\protect\citeauthoryear{{Gregory}}{{Gregory}}{2005}]{Gregory05}
{Gregory} P.~C.,  2005, {Bayesian Logical Data Analysis for the Physical
  Sciences: A Comparative Approach with `Mathematica' Support}.
Cambridge University Press

\bibitem[\protect\citeauthoryear{{Hauschildt}, {Allard}  \&
  {Baron}}{{Hauschildt} et~al.}{1999}]{Hauschildt99}
{Hauschildt} P.~H.,  {Allard} F.,   {Baron} E.,  1999, \mn@doi [\apj]
  {10.1086/306745}, \href {http://adsabs.harvard.edu/abs/1999ApJ...512..377H}
  {512, 377}

\bibitem[\protect\citeauthoryear{{Hessels} et~al.,}{{Hessels}
  et~al.}{2011}]{Hessels11}
{Hessels} J.~W.~T.,  et~al., 2011, in {Burgay} M.,  {D'Amico} N.,  {Esposito}
  P.,  {Pellizzoni} A.,   {Possenti} A.,  eds,  American Institute of Physics
  Conference Series Vol. 1357, American Institute of Physics Conference Series.
  pp 40--43 (\mn@eprint {arXiv} {1101.1742}), \mn@doi{10.1063/1.3615072}

\bibitem[\protect\citeauthoryear{{Hill}, {Watson}, {Shahbaz}, {Steeghs}  \&
  {Dhillon}}{{Hill} et~al.}{2014}]{Hill14}
{Hill} C.~A.,  {Watson} C.~A.,  {Shahbaz} T.,  {Steeghs} D.,   {Dhillon} V.~S.,
   2014, \mn@doi [\mnras] {10.1093/mnras/stu1460}, \href
  {http://adsabs.harvard.edu/abs/2014MNRAS.444..192H} {444, 192}

\bibitem[\protect\citeauthoryear{{Kaplan}, {Bhalerao}, {van Kerkwijk},
  {Koester}, {Kulkarni}  \& {Stovall}}{{Kaplan} et~al.}{2013}]{Kaplan13}
{Kaplan} D.~L.,  {Bhalerao} V.~B.,  {van Kerkwijk} M.~H.,  {Koester} D.,
  {Kulkarni} S.~R.,   {Stovall} K.,  2013, \mn@doi [\apj]
  {10.1088/0004-637X/765/2/158}, \href
  {http://adsabs.harvard.edu/abs/2013ApJ...765..158K} {765, 158}

\bibitem[\protect\citeauthoryear{{Kiziltan}, {Kottas}, {De Yoreo}  \&
  {Thorsett}}{{Kiziltan} et~al.}{2013}]{Kiziltan13}
{Kiziltan} B.,  {Kottas} A.,  {De Yoreo} M.,   {Thorsett} S.~E.,  2013, \mn@doi
  [\apj] {10.1088/0004-637X/778/1/66}, \href
  {http://adsabs.harvard.edu/abs/2013ApJ...778...66K} {778, 66}

\bibitem[\protect\citeauthoryear{{Lattimer} \& {Prakash}}{{Lattimer} \&
  {Prakash}}{2004}]{Lattimer04}
{Lattimer} J.~M.,  {Prakash} M.,  2004, \mn@doi [Science]
  {10.1126/science.1090720}, \href
  {http://adsabs.harvard.edu/abs/2004Sci...304..536L} {304, 536}

\bibitem[\protect\citeauthoryear{{Li}, {Halpern}  \& {Thorstensen}}{{Li}
  et~al.}{2014}]{Li14}
{Li} M.,  {Halpern} J.~P.,   {Thorstensen} J.~R.,  2014, \mn@doi [\apj]
  {10.1088/0004-637X/795/2/115}, \href
  {http://adsabs.harvard.edu/abs/2014ApJ...795..115L} {795, 115}

\bibitem[\protect\citeauthoryear{{Li}, {Kong}, {Hou}, {Mao}, {Strader},
  {Chomiuk}  \& {Tremou}}{{Li} et~al.}{2016}]{Li16}
{Li} K.-L.,  {Kong} A.~K.~H.,  {Hou} X.,  {Mao} J.,  {Strader} J.,  {Chomiuk}
  L.,   {Tremou} E.,  2016, \mn@doi [\apj] {10.3847/1538-4357/833/2/143}, \href
  {http://adsabs.harvard.edu/abs/2016ApJ...833..143L} {833, 143}

\bibitem[\protect\citeauthoryear{{Linares}}{{Linares}}{2014}]{Linares14}
{Linares} M.,  2014, \mn@doi [\apj] {10.1088/0004-637X/795/1/72}, \href
  {http://adsabs.harvard.edu/abs/2014ApJ...795...72L} {795, 72}

\bibitem[\protect\citeauthoryear{{Linares}, {Miles-P{\'a}ez},
  {Rodr{\'{\i}}guez-Gil}, {Shahbaz}, {Casares}, {Fari{\~n}a}  \&
  {Karjalainen}}{{Linares} et~al.}{2016}]{Linares16}
{Linares} M.,  {Miles-P{\'a}ez} P.,  {Rodr{\'{\i}}guez-Gil} P.,  {Shahbaz} T.,
  {Casares} J.,  {Fari{\~n}a} C.,   {Karjalainen} R.,  2016, preprint, \href
  {http://adsabs.harvard.edu/abs/2016arXiv160902232L} {} (\mn@eprint {arXiv}
  {1609.02232})

\bibitem[\protect\citeauthoryear{{Lucy}}{{Lucy}}{1967}]{Lucy67}
{Lucy} L.~B.,  1967, \zap, \href
  {http://adsabs.harvard.edu/abs/1967ZA.....65...89L} {65, 89}

\bibitem[\protect\citeauthoryear{{Marsh}, {Robinson}  \& {Wood}}{{Marsh}
  et~al.}{1994}]{Marsh94}
{Marsh} T.~R.,  {Robinson} E.~L.,   {Wood} J.~H.,  1994, \mn@doi [\mnras]
  {10.1093/mnras/266.1.137}, \href
  {http://adsabs.harvard.edu/abs/1994MNRAS.266..137M} {266, 137}

\bibitem[\protect\citeauthoryear{{Orosz} \& {Hauschildt}}{{Orosz} \&
  {Hauschildt}}{2000}]{Orosz00}
{Orosz} J.~A.,  {Hauschildt} P.~H.,  2000, \aap, \href
  {http://adsabs.harvard.edu/abs/2000A%26A...364..265O} {364, 265}

\bibitem[\protect\citeauthoryear{{Orosz} \& {van Kerkwijk}}{{Orosz} \& {van
  Kerkwijk}}{2003}]{Orosz03}
{Orosz} J.~A.,  {van Kerkwijk} M.~H.,  2003, \mn@doi [\aap]
  {10.1051/0004-6361:20021468}, \href
  {http://adsabs.harvard.edu/abs/2003A%26A...397..237O} {397, 237}

\bibitem[\protect\citeauthoryear{{{\"O}zel} \& {Freire}}{{{\"O}zel} \&
  {Freire}}{2016}]{Ozel16}
{{\"O}zel} F.,  {Freire} P.,  2016, \mn@doi [\araa]
  {10.1146/annurev-astro-081915-023322}, \href
  {http://adsabs.harvard.edu/abs/2016ARA%26A..54..401O} {54, 401}

\bibitem[\protect\citeauthoryear{{{\"O}zel}, {Psaltis}, {Narayan}  \& {Santos
  Villarreal}}{{{\"O}zel} et~al.}{2012}]{Ozel12}
{{\"O}zel} F.,  {Psaltis} D.,  {Narayan} R.,   {Santos Villarreal} A.,  2012,
  \mn@doi [\apj] {10.1088/0004-637X/757/1/55}, \href
  {http://adsabs.harvard.edu/abs/2012ApJ...757...55O} {757, 55}

\bibitem[\protect\citeauthoryear{{Papitto} et~al.,}{{Papitto}
  et~al.}{2013}]{Papitto13}
{Papitto} A.,  et~al., 2013, \mn@doi [\nat] {10.1038/nature12470}, \href
  {http://adsabs.harvard.edu/abs/2013Natur.501..517P} {501, 517}

\bibitem[\protect\citeauthoryear{{Pecaut} \& {Mamajek}}{{Pecaut} \&
  {Mamajek}}{2013}]{Pecaut13}
{Pecaut} M.~J.,  {Mamajek} E.~E.,  2013, \mn@doi [\apjs]
  {10.1088/0067-0049/208/1/9}, \href
  {http://adsabs.harvard.edu/abs/2013ApJS..208....9P} {208, 9}

\bibitem[\protect\citeauthoryear{{Phillips}, {Shahbaz}  \&
  {Podsiadlowski}}{{Phillips} et~al.}{1999}]{Phillips99}
{Phillips} S.~N.,  {Shahbaz} T.,   {Podsiadlowski} P.,  1999, \mn@doi [\mnras]
  {10.1046/j.1365-8711.1999.02357.x}, \href
  {http://adsabs.harvard.edu/abs/1999MNRAS.304..839P} {304, 839}

\bibitem[\protect\citeauthoryear{{Roberts}}{{Roberts}}{2013}]{Roberts13}
{Roberts} M.~S.~E.,  2013, in {van Leeuwen} J.,  ed.,  IAU Symposium Vol. 291,
  Neutron Stars and Pulsars: Challenges and Opportunities after 80 years. pp
  127--132 (\mn@eprint {arXiv} {1210.6903}), \mn@doi{10.1017/S174392131202337X}

\bibitem[\protect\citeauthoryear{{Romani} \& {Sanchez}}{{Romani} \&
  {Sanchez}}{2016}]{Romani16}
{Romani} R.~W.,  {Sanchez} N.,  2016, \mn@doi [\apj]
  {10.3847/0004-637X/828/1/7}, \href
  {http://adsabs.harvard.edu/abs/2016ApJ...828....7R} {828, 7}

\bibitem[\protect\citeauthoryear{{Romani}, {Filippenko}  \& {Cenko}}{{Romani}
  et~al.}{2015}]{Romani15}
{Romani} R.~W.,  {Filippenko} A.~V.,   {Cenko} S.~B.,  2015, \mn@doi [\apj]
  {10.1088/0004-637X/804/2/115}, \href
  {http://adsabs.harvard.edu/abs/2015ApJ...804..115R} {804, 115}

\bibitem[\protect\citeauthoryear{{Schroeder} \& {Halpern}}{{Schroeder} \&
  {Halpern}}{2014}]{Schroeder14}
{Schroeder} J.,  {Halpern} J.,  2014, \mn@doi [\apj]
  {10.1088/0004-637X/793/2/78}, \href
  {http://adsabs.harvard.edu/abs/2014ApJ...793...78S} {793, 78}

\bibitem[\protect\citeauthoryear{{Shahbaz}}{{Shahbaz}}{1998}]{Shahbaz98}
{Shahbaz} T.,  1998, \mn@doi [\mnras] {10.1046/j.1365-8711.1998.01618.x}, \href
  {http://adsabs.harvard.edu/abs/1998MNRAS.298..153S} {298, 153}

\bibitem[\protect\citeauthoryear{{Shahbaz}}{{Shahbaz}}{2003}]{Shahbaz03b}
{Shahbaz} T.,  2003, \mn@doi [\mnras] {10.1046/j.1365-8711.2003.06258.x}, \href
  {http://adsabs.harvard.edu/abs/2003MNRAS.339.1031S} {339, 1031}

\bibitem[\protect\citeauthoryear{{Shahbaz}, {Groot}, {Phillips}, {Casares},
  {Charles}  \& {van Paradijs}}{{Shahbaz} et~al.}{2000}]{Shahbaz00}
{Shahbaz} T.,  {Groot} P.,  {Phillips} S.~N.,  {Casares} J.,  {Charles} P.~A.,
   {van Paradijs} J.,  2000, \mn@doi [\mnras]
  {10.1046/j.1365-8711.2000.03341.x}, \href
  {http://adsabs.harvard.edu/abs/2000MNRAS.314..747S} {314, 747}

\bibitem[\protect\citeauthoryear{{Shahbaz}, {Zurita}, {Casares}, {Dubus},
  {Charles}, {Wagner}  \& {Ryan}}{{Shahbaz} et~al.}{2003}]{Shahbaz03a}
{Shahbaz} T.,  {Zurita} C.,  {Casares} J.,  {Dubus} G.,  {Charles} P.~A.,
  {Wagner} R.~M.,   {Ryan} E.,  2003, \mn@doi [\apj] {10.1086/346001}, \href
  {http://adsabs.harvard.edu/abs/2003ApJ...585..443S} {585, 443}

\bibitem[\protect\citeauthoryear{{Shahbaz}, {Casares}, {Watson}, {Charles},
  {Hynes}, {Shih}  \& {Steeghs}}{{Shahbaz} et~al.}{2004}]{Shahbaz04}
{Shahbaz} T.,  {Casares} J.,  {Watson} C.~A.,  {Charles} P.~A.,  {Hynes} R.~I.,
   {Shih} S.~C.,   {Steeghs} D.,  2004, \mn@doi [\apjl] {10.1086/426504}, \href
  {http://adsabs.harvard.edu/abs/2004ApJ...616L.123S} {616, L123}

\bibitem[\protect\citeauthoryear{{Shahbaz}, {Watson}  \& {Dhillon}}{{Shahbaz}
  et~al.}{2014}]{Shahbaz14}
{Shahbaz} T.,  {Watson} C.~A.,   {Dhillon} V.~S.,  2014, \mn@doi [\mnras]
  {10.1093/mnras/stu267}, \href
  {http://adsabs.harvard.edu/abs/2014MNRAS.440..504S} {440, 504}

\bibitem[\protect\citeauthoryear{{Stappers}, {van Kerkwijk}, {Bell}  \&
  {Kulkarni}}{{Stappers} et~al.}{2001}]{Stappers01}
{Stappers} B.~W.,  {van Kerkwijk} M.~H.,  {Bell} J.~F.,   {Kulkarni} S.~R.,
  2001, \mn@doi [\apjl] {10.1086/319106}, \href
  {http://adsabs.harvard.edu/abs/2001ApJ...548L.183S} {548, L183}

\bibitem[\protect\citeauthoryear{Storn \& Price}{Storn \&
  Price}{1995}]{Storn95}
Storn R.,  Price K.,  1995, Differential Evolution - A simple and efficient
  adaptive scheme for global optimization over continuous spaces

\bibitem[\protect\citeauthoryear{{Strader}, {Chomiuk}, {Sonbas}, {Sokolovsky},
  {Sand}, {Moskvitin}  \& {Cheung}}{{Strader} et~al.}{2014}]{Strader14}
{Strader} J.,  {Chomiuk} L.,  {Sonbas} E.,  {Sokolovsky} K.,  {Sand} D.~J.,
  {Moskvitin} A.~S.,   {Cheung} C.~C.,  2014, \mn@doi [\apjl]
  {10.1088/2041-8205/788/2/L27}, \href
  {http://adsabs.harvard.edu/abs/2014ApJ...788L..27S} {788, L27}

\bibitem[\protect\citeauthoryear{{Tang} et~al.,}{{Tang} et~al.}{2014}]{Tang14}
{Tang} S.,  et~al., 2014, \mn@doi [\apjl] {10.1088/2041-8205/791/1/L5}, \href
  {http://adsabs.harvard.edu/abs/2014ApJ...791L...5T} {791, L5}

\bibitem[\protect\citeauthoryear{Vrugt}{Vrugt}{2016}]{Vrugt2016}
Vrugt J.~A.,  2016, \mn@doi [Environmental Modelling & Software]
  {http://dx.doi.org/10.1016/j.envsoft.2015.08.013}, 75, 273

\bibitem[\protect\citeauthoryear{{Wang}, {Archibald}, {Thorstensen}, {Kaspi},
  {Lorimer}, {Stairs}  \& {Ransom}}{{Wang} et~al.}{2009}]{Wang09}
{Wang} Z.,  {Archibald} A.~M.,  {Thorstensen} J.~R.,  {Kaspi} V.~M.,  {Lorimer}
  D.~R.,  {Stairs} I.,   {Ransom} S.~M.,  2009, \mn@doi [\apj]
  {10.1088/0004-637X/703/2/2017}, \href
  {http://adsabs.harvard.edu/abs/2009ApJ...703.2017W} {703, 2017}

\bibitem[\protect\citeauthoryear{{Welsh}, {Horne}  \& {Gomer}}{{Welsh}
  et~al.}{1995}]{Welsh95}
{Welsh} W.~F.,  {Horne} K.,   {Gomer} R.,  1995, \mn@doi [\mnras]
  {10.1093/mnras/275.3.649}, \href
  {http://adsabs.harvard.edu/abs/1995MNRAS.275..649W} {275, 649}

\bibitem[\protect\citeauthoryear{{van Kerkwijk}, {Breton}  \& {Kulkarni}}{{van
  Kerkwijk} et~al.}{2011}]{Kerkwijk11}
{van Kerkwijk} M.~H.,  {Breton} R.~P.,   {Kulkarni} S.~R.,  2011, \mn@doi
  [\apj] {10.1088/0004-637X/728/2/95}, \href
  {http://adsabs.harvard.edu/abs/2011ApJ...728...95V} {728, 95}

\bibitem[\protect\citeauthoryear{{van Staden} \& {Antoniadis}}{{van Staden} \&
  {Antoniadis}}{2016}]{Staden16}
{van Staden} A.~D.,  {Antoniadis} J.,  2016, \mn@doi [\apjl]
  {10.3847/2041-8213/833/1/L12}, \href
  {http://adsabs.harvard.edu/abs/2016ApJ...833L..12V} {833, L12}

\makeatother
\end{thebibliography}
\end{document}